\documentclass[12pt]{article}
\pdfoutput=1
\usepackage{subfigure}
\usepackage{amssymb,amsmath}
\usepackage{graphicx}
\usepackage{color}
\usepackage[colorlinks=true
,urlcolor=blue
,citecolor=blue
,linkcolor=blue
,pagecolor=blue
,linktocpage=true
,pdfproducer=medialab
]{hyperref}
\usepackage[a4paper]{geometry}

\usepackage{placeins}
\usepackage{slashed}

\makeatletter \renewcommand{\@dotsep}{10000} \makeatother

\mathchardef\mhyphen="2D

\newcommand{\beq}{\begin{equation}}
\newcommand{\eeq}{\end{equation}}
\newcommand{\bea}{\begin{eqnarray}}
\newcommand{\eea}{\end{eqnarray}}

\numberwithin{equation}{section}

\begin{document}

\begin{titlepage}
\pagestyle{empty}

\vspace*{0.2in}
\begin{center}
{\Large\bf    Associated Production of $Z_{H}$ and $T$ in the Littlest Higgs Model at High Energy Linear $e^{-}e^{+}$ Collider}\\
\vspace{1cm}
{\large Zerrin Kirca\footnote{E-mail: zkirca@uludag.edu.tr}}
\vspace{0.5cm}

{\it
Department of Physics, Uluda\~{g} University, TR16059 Bursa, Turkey
}

\end{center}

\vspace{0.5cm}
\begin{abstract}
\noindent
In this work we study associated production of heavy Z boson and Heavy top quark in the framework of Littlest Higgs model at $e^{-}e^{+}$ colliders considering parameter space  allowed by the electroweak precision measurements. According to the free parameters of the model the possibility of detecting of new heavy particles at $\sqrt{s}$=3000 GeV and yearly integrated luminosity of $L$=500$fb^{-1}$ are discussed. We find that in a narrow range of the parameter space, $s/s^{\prime}$=0.8/0.6,0.7, 0.4 $\leqslant x_{L}\leqslant$ 0.6 and $f\lesssim$ 1060 GeV, a statical significance of 5$\sigma$ can be achieved. We also discuss constraints on masses of heavy top quark and heavy $Z$ boson together with the mixing parameters $s$ and $x_L$ at $\sqrt{s}$=3000 GeV.

\end{abstract}

\end{titlepage}

\section{Introduction}
In spite of all experimental tests that Standard Model (SM) has passed successfully, it contains many theoretical problems. Therefore SM is assumed to be an effective theory that is valid up to a certain energy scale. Solutions of these shortcomings are proposed by new physics beyond the SM which should exist at TeV scale. Hierarchy problem is the main motivation for the existence of new physics at the TeV scale. It is the problem about Higgs scalar receiving quadratically divergent loop contribution to its mass through its coupling to SM particles; the top quark, the weak gauge bosons and its self quadratic term and the mass term requires fine-tuning of huge quantum corrections cancelling each other. In order to solve the hierarchy problem, many extended theories are proposed beyond the SM such as supersymmetry, extra dimensional model, left-right twin Higgs model and little Higgs  models.

The Little Higgs models provide a solution to the hierarchy problem by regarding the SM Higgs particle as a Pseudo-Goldstone boson of a new global symmetry group~\cite{Arkani-Hamed1, Arkani-Hamed2, Arkani-Hamed3, D.E. Kaplan4, Low5}.  The models using collective symmetry breaking can be divided into two groups which provide a natural way giving the Higgs boson mass: product group models and simple group models. Their main difference is the choice of the new global symmetry. As a result of collective symmetry breaking, Little Higgs models contain, in addition to the SM, new heavy particles whose contributions cancel the loop quadratic divergences in the Higgs mass. This cancellation occurs between  particles which have the same quantum numbers as the SM. There are several studies in literature on phenomenology of Little Higgs Models ~\cite{Arkani-Hamed6, Han7, Schmalz8}. Also the models which have different theory space get constraints by electroweak precision measurement data ~\cite{Csaki9, Csaki10, Hewett11, Chen12, Kilian13, Meade14}.

The minimal version of the Little Higgs Models is  "Littlest Higgs Model" (LH)~\cite{Arkani-Hamed6}. It is a product group model which is based on $SU(5)/SO(5)$ non linear sigma model. The global $SU(5)$ symmetry containing a gauged subgroup $\left[SU(2)_{1}\times U(1)_{1} \right]\times \left[SU(2)_{2}\times U(1)_{2} \right] $ is broken to a global $SO(5)$ symmetry at scale $\Lambda$ by a vacuum expectation value of order of $f$. The gauged symmetry $\left[SU(2)_{1}\times U(1)_{1} \right]\times \left[SU(2)_{2}\times U(1)_{2} \right] $ is spontaneously broken into the $SU(2)_{W}\times U(1)_{Y}$ , identified as the SM gauge group. This symmetry breaking results in fourteen Goldstone bosons which are parametrized by a non linear sigma field. Four of them are eaten by the heavy gauge bosons ($Z_{H}, W^{\mp}_{H}, A_{H}$) associated with the broken symmetry. After these gauge bosons obtain their masses, the remaining ten Goldstone bosons will come the complex doublet ($h^{+},h^{0}$) and complex scalar triplet ($\phi^{0}, \phi^{P}, \phi^{+}$ and $\phi^{++}$) of the electroweak gauge group which is identified as the SM Higgs and complex triplet which gains a mass of the scale $f$, respectively. Also a new vector-like top quark ($T$) which is the heavy partner of the SM top quark is taken into account to cancel the quadratically divergent contribution from Higgs coupling to the SM top quark. The new heavy particles proposed by the LH interact with each other and with the SM particles. 

The heavy vector like top quark is commonly predicted by many extensions of the SM and limits on the existence of the new particles have been studied in the literature as model dependent and independent~\cite{Buras15, Saavadre16, Vignaroli17, Okada18, Matsedoskyi19, Gripaios20, Liu21, CHan22}. LHC collaborations have performed the searches for the heavy top quark using first run of LHC. the heavy top quark with mass less than 700 GeV has been excluded by ATLAS ~\cite{ATLAS23} and CMS ~\cite{CMS24}.

In this work, the  productions of the new heavy gauge boson $Z_{H}$ and  heavy top quark $T$ via $e^{-}e^{+} \rightarrow Z_{H}\overline{T}t$ in the Littlest Higgs model are examined. High energy linear $e^{-}e^{+}$ colliders which have the clean environment are of particular importance to study production of the new heavy particles predicted by the Little Higgs Models ~\cite{LC25}. In literature, there are several studies which have analyzed the Little Higgs models in $e^{-}e^{+}$ which can be transformed to $e\gamma$~ and $\gamma\gamma$ and LHC considering heavy top quark and heavy Z gauge boson ~\cite{Yue26, Liu27, Maldonado28, Maldonado29, Harigaya30, Yang31, Yang32, Aranda33}.  

The rest of this paper is organized as follows: In section II we give a brief review of 
general features of the Littlest Higgs model. In Sec. III we will discuss associated production of heavy Z boson and heavy top quark in $e^{-}e^{+}$ collisions according to free parameters of the model. Finally, the conclusions are presented in Sec. IV.

\section{The Model}

Littlest Higgs model is a non linear sigma model based on $SU(5)$ global symmetry 
which contains a gauged $[SU(2)\times U(1)]_{1}\times[SU(2)\times U(1)]_{2}$ subgroup. 
The global $SU(5)$ is broken into $SO(5)$ via vacuum expectation value of the sigma field at the $f$ scale $\sim$ 1 TeV~\cite{Arkani-Hamed6},
\begin{equation} 
 \Sigma = e^{\frac{i\Pi}{f}}\Sigma_{0} e^{\frac{i\Pi^{{T}}}{f}}.
 \label{eq1}
\end{equation}
The symmetry breaking vacuum expectation value is chosen as

\begin{equation}
 \Sigma_{0} =\left(\begin{array}{ccc} 
0 & 0 &I\\ 
0 & 1 & 0 \\ 
I & 0 & 0 \end{array}\right) 
\end{equation}
where $I$ is the $2\times2$ identity matrix. As a result of the symmetry breaking there will be fourteen massless Goldstone bosons. Four of these will be removed by the Higgs mechanism when the $[SU(2)\times U(1)]^{2}$ is broken to its diagonal subgroup which is identified as the $[SU(2)\times U(1)]$. The remaining Goldstone bosons are parametrized by a non-linear sigma model field where the Goldstone boson matrix $\Pi$ is

\begin{equation}
\Pi =\left(\begin{array}{ccc} 
 & h^{\dagger}/\sqrt{2} & \phi^{\dagger} \\ 
h/\sqrt{2} &  & h^{\ast}/\sqrt{2} \\ 
\phi & h^{T}/\sqrt{2} &  \end{array}\right) 
\end{equation}
where $h$ is a complex doublet, $h=(h^{+},h^{0})$ and $\phi$ is a complex electroweak doublet which is form

\begin{equation*}
\phi =\left(\begin{array}{cc} 
\phi^{++} & \phi^{+}/\sqrt{2}\\ 
\phi^{+}/\sqrt{2} & \phi^{0}  
\end{array}\right).
\end{equation*}

The kinetic term for the $\Sigma$ field can be written as~\cite{Arkani-Hamed6}
\begin{equation}
\mathcal{L}_{kin}=\frac{f^{2}}{8}Tr\left\lbrace D_{\mu}\Sigma(D^{\mu}\Sigma)^{\dagger}\right\rbrace ,
\end{equation}
where~\cite{Arkani-Hamed6}
\begin{equation} 
D_{\mu}\Sigma=\partial_{\mu}\Sigma-i\sum_{i=1}^{2}\left[ g_{i}(W_{\mu i}\Sigma%
+\Sigma W_{\mu i}^{{T}})+
g_{i}^{\prime}(B_{\mu i}\Sigma+\Sigma B_{\mu i}^{{T}})\right] 
\end{equation}
$W_{\mu i}$ and $B_{\mu i}$ are the $SU(2)$ and $U(1)$ gauge fields, respectively, and $g_{i}$ and $g_{i}^{\prime}$ are the respective gauge couplings. The orthogonal combinations of gauge bosons are identified as SM gauge bosons and the heavy gauge bosons mass eigenstates are given by~\cite{Arkani-Hamed6}
\begin{equation}
\begin{array}{ccc}
W&=& s W_1 + c W_2 \qquad  W' = -c W_1 + s W_2 \\ 
B&=& s' B_1 + c' B_2 \qquad B' = -c' B_1 + s' B_2~. 
\end{array}
\end{equation} 
($c, c^{\prime}, s, s^{\prime} $) denote the cosine and sine of mixing angles, respectively and they are described by the gauge couplings of the $SU(2)_L \times U(1)_Y$ groups; $g=g_{1}s=g_{2}c$ and $g'=g'_{1}s'=g'_{2}c'$. The $g$ and $g^\prime$ are the SM couplings,
\begin{equation}
\frac 1 {g^2}=\frac 1{g_1^2}+\frac 1{g_2^2},~~~~~~~ \frac 1 
{{g'}^2}=\frac 1{{g_1'}^2}+\frac 1{{g_2'}^2}~. 
\end{equation} 
At the scale $f$, the SM gauge bosons $W$ and $B$ remain massless ($m_{W}$ and $m_{B}$) while the heavy gauge bosons ($m_{W^\prime}$ and $m_{B^\prime}$) acquire masses as $m_{W^\prime}=gf/2sc$ and $m_{B^\prime}=g{^\prime}f/2 \sqrt{5}s^{\prime} c^{\prime}$. In the scalar sector of the LH, Coleman-Weinberg potential triggers the electroweak symmetry breaking then $h$ and $\phi$  acquire vacuum expectation values:$\left\langle h^{0}\right\rangle=v/\sqrt{2}$ and $\left\langle \phi^{0} \right\rangle=v'$. By the electroweak symmetry breaking there is a mixing between SM gauge bosons and heavy gauge bosons due to vev of $h$ and $\phi$. The final mass eigenstates of charged and neutral gauge bosons at order of $v^2/f^2$ are given by\cite{Buras15}  
\begin{equation}
\nonumber
M_{A_L}^2=0,~~~~~~~M_{W_L^\pm}^2=m_{W}^2\left(1-\frac{v^2}{f^2}\left(\frac{1}{6}+\frac{1}{4}\left(c^2-s^2 \right)^2\right)\right) 
\end{equation} 

\begin{equation}
M_{Z_L}^2=m_{Z}^2\left(1-\frac{v^2}{f^2}\left(\frac{1}{6}+\frac{1}{4}\left(c^2-s^2 \right)^2+\frac{5}{4}\left(c'^2-s'^2 \right)^2\right)\right)
\end{equation}

\begin{equation}
M_{A_H}^2=m_{Z}^2 s_{W}^2\left(\frac{f^2}{5{s'}^2{c'}^2{v}^2}-1\right)~~~~~M_{Z_H}^2=m_{W}^2\left(\frac{f^2}{{s}^2{c}^2{v}^2}-1\right)\approx M_{W_H^\pm}^2
\end{equation} 
where $L$ and $H$ indices denote $light$ and $heavy$ gauge bosons respectively. $A_L$ remains massless while $W^{\mp}_L$ and $Z_L$ masses get correction of order of $v^2/f^2$. Here $s_w$ and $c_w$ are the usual mixing angles, $s_w=\frac{g^{\prime}}{\sqrt{g^2+{g^\prime}^2}}$ and $c_w=\frac{g}{\sqrt{g^2+{g^\prime}^2}}$, $m_W$ and $m_Z$ are the SM gauge boson masses and $v= 246$ GeV.

\begin{figure}[h!]\hspace{-1.5cm}
\subfigure{\includegraphics[scale=0.5]{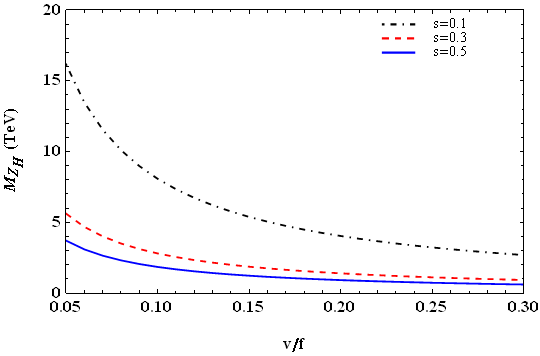}}%
\subfigure{\includegraphics[scale=0.49]{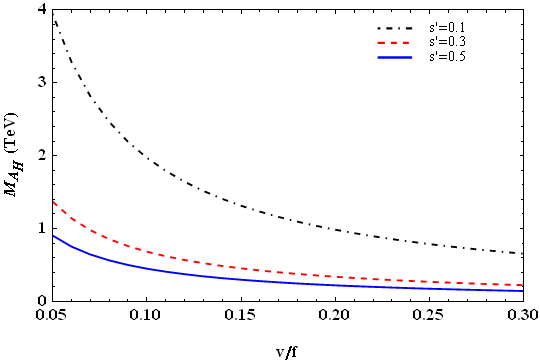}}
\caption{The masses of the $Z_H$($left$) and $A_H$ ($right$) as functions of $v/f$ for different values of $s$ and $s'$=0.1 (dot-dashed line), 0.3 (dashed line) and 0.5 (solid line).}
\label{MZH-MAH-fig}
\end{figure}

As can be seen in Fig. 1, the mass of the $Z_H$($A_H$) mainly depends on the free parameters of the model ($s$($s'$) and $f$). For $v/f>$0.081, the mass of the heavy photon is lighter than 500 GeV in the limits on Littlest Higgs model which is constrained by the electroweak precision measurement while the mass of the heavy Z boson is lighter than 2 TeV as can be seen in Table 1 and Fig 1. So $A_H$ is much lighter than $Z_H$, $M_{Z_H}\simeq 4M_{A_H}$. So the future colliders should get the first signal of $A_H$. Decay channels of    the $Z_H$ have been analyzed to test the Little Higgs Models in ~\cite{Yue26, Maldonado28, Maldonado29, Aranda33}.

In the LH model the couplings between fermions and gauge bosons are written in the form $i\gamma_{\mu}$($g_{V}$+$g_{A}\gamma_{5}$). Vector and axial couplings depend on free parameters of the model: $s$, $s^{^\prime}$, $f$ and $x_{L}$. Relevant couplings are given in Table 2. The decay width of the heavy bosons into fermion pairs can be written as $\Gamma_{V_i}=\frac{C}{24\pi}(g_{V}^2+g_{A}^2)M_{V_i}$ where $C$ is 3 for quarks and $C$ is 1 for leptons. Then total decay widths of the heavy gauge bosons as functions of the masses are given as ~\cite{Yue26}

\begin{equation*}
\Gamma_{Z_H}=\frac{g^2 M_{Z_H}(193-388s^2+196s^4)}{768\pi s^2(1-s^2)}
\end{equation*} 

\begin{equation}
\Gamma_{W_H}=\frac{g^2 M_{W_H}(97-196s^2+100s^4)}{384\pi s^2(1-s^2)}
\end{equation}

\begin{equation*}
\Gamma_{A_H}=\frac{g'^2 M_{A_H}(21-70s'^2+59s'^4)}{48\pi s'^2(1-s'^2)}
\end{equation*}
The masses and the total decay widths of heavy gauge bosons are given in Table 1 for $v/f=0.1$.

\begin{table}[h!]
\centering
\begin{tabular}{|c|c|c|c|c|c|}
\hline 
$s$/$s^{\prime}$ & $M_{Z_H}$, $M_{W_H}$ (GeV)&$M_{A_H}$ (GeV)&$\Gamma_{Z_H}$(GeV) &$\Gamma_{W_H}$ (GeV) &$\Gamma_{A_H}$ (GeV) \\
\hline
0.1/0.1 & 8082  & 1973.2& 27094.2 & 27231.8 &  3430.5\\
\hline 
0.2/0.3 & 4103.3& 684.8& 3333.5  & 3349.4 &  107.6\\
\hline 
0.5/0.5 & 1855.5& 451.4 & 187.9 & 188.4 &  14.7\\
\hline 
0.7/0.6 &1606.8& 406.8 & 56.3  &56.4 &5.2  \\
\hline 
0.8/0.7 &1673.5 & 390.4&31.8  & 31.9 & 1.1 \\
\hline 
0.9/0.9 &2048.5 &498.7 & 17.1 &16.9  &  8.3\\
\hline 
\end{tabular}
\caption{Total decay widths and masses of heavy gauge bosons with respect to LH parameters}
\end{table}

In the LH model, fermion sector introduces a new pair of heavy vector-like fermions which couple to the Higgs in order to cancel the quadratic divergences to the Higgs mass due to the SM top quark. For the first two generations there is no need to introduce heavy partners because their Yukawa couplings are much smaller than the top quark and do not have an important contribution to the quadratic divergence of the Higgs mass. $\tilde t$ and $\tilde t'^{c}$ are new top quarks proposed by the LH model. Top sector Lagrangian can be written as~\cite{Arkani-Hamed6}
\begin{equation} 
\mathcal{L}_{t}=\frac{1}{2}\lambda_{1}f\epsilon_{ijk}\epsilon_{x,y}\chi_{i}\Sigma_{j,x}\Sigma_{k,y}{u'}^c_{3}+\lambda_{2}f\tilde{t}\tilde{t'}^c+h.c.
\end{equation} 
where $\chi$ is the row vector,$\chi_{i}=(b_{3},t_{3},\tilde{t})$ and ${u'}_{3}$ is the right-handed top quark of the SM. The indices $i, j, k$ are summed over 1, 2, 3 and $x,y$ over 4, 5. The first and second term give the coupling of the Higgs boson to fermions and the mass term of the heavy top quark, respectively. 
If $\mathcal{L}_{t}$ is expanded around the vacuum expectation value $\Sigma$ field, we obtain~\cite{Arkani-Hamed6} 
\begin{equation} 
\mathcal{L}_{t}=\lambda_{2}f\tilde{t}\tilde{t'}^c+{\lambda_{1}}f{\tilde t}{u'}^c_{3}-\frac{\lambda_{1}}{f}{\tilde t}h^{\dagger}h{u'}^c_{3}-i{\lambda_{1}}{\sqrt{2}}q_{3}h^{\dagger}{u'}^c_{3}+h.c.+...
\end{equation}
where $q_{3}$ is a row vector, $q_{3}=(b_3,t_3)$. We have omitted the contribution of ${\phi}$ boson in Eq (2.12). Before electroweak symmetry breaking, the mass eigenstates which is defining mix of $\tilde{t}^c$ and ${u'}^c_{3}$ are given by
\begin{equation}
t^c_{R}={u}^c_{3}=\frac{\lambda_{2}{u'}^c_{3}-\lambda_{1}\tilde{t'}^c}{\sqrt{\lambda_{1}^2+\lambda_{2}^2}},\qquad T^c_{R}={\tilde{t}^c}=\frac{\lambda_{2}\tilde{t'}^c+\lambda_{1}{u'}^c_{3}}{\sqrt{\lambda_{1}^2+\lambda_{2}^2}}~~.
\end{equation}
At the scale $f$, $t_{L}=t_{3}$ which is identified as SM top quark is massless  while the mass of heavy top partner ($\tilde{t}$) is defined by
\begin{equation}
m_{\tilde{t}}=f\sqrt{\lambda_{1}^2+\lambda_{2}^2}~~.
\end{equation} 
After electroweak symmetry breaking, the mass eigenstates of the top quark and its heavy partner $T$ are given by
\begin{equation}
 t_{L}=c_{L}t_{3}-s_{L}\tilde{t},\qquad t^c_{R}=c_{R}{u'}^c_{3}-s_{R}\tilde{t'}^c
 \end{equation}
\begin{equation}
 T_{L}=s_{L}t_{3}+c_{L}\tilde{t},\qquad T^c_{R}=s_{R}{u'}^c_{3}+c_{R}\tilde{t'}^c
\end{equation}
The mixing between the right-handed top quark and heavy top quark can be written shortly~\cite{Buras15}   
\begin{equation}
x_{L}=\frac{\lambda_{1}^2}{\lambda_{1}^{2}+\lambda_{2}^{2}}
\end{equation}
when $x_{L}=\lambda_{1}=$0, there is no mixing between right handed top quark and heavy top quark. If $x_{L}=$1 ($\lambda_{2}=$0) the mass eigenstate of $\tilde{t}^c$ is equal to the right handed top quark, $u_{3}'$ and the mixing angles are 
\begin{equation*}
c_ {L}=1-\frac{v^{2}}{f^{2}}\frac{x_{L}^2}{2},~~~~~~s_ {L}=x_{L}\frac{v}{f}
\left[1+\frac{v^2}{f^2}\left(-\frac{5}{6}+\frac{fv'}{v^2}+x_{L}\left(2-\frac{3}{2}x_{L}\right) \right)  \right] 
\end{equation*}
\begin{equation*}
s_ {R}=\sqrt{x_{L}}\left[1-\frac{v^2}{2f^2}\left(1-3x_{L}+2x_{L}^2\right)\right],~~~~~~c_ {R}=\sqrt{1-x_{L}}\left[1+\frac{v^2}{f^2}\left(\frac{1}{2}x_{L}-x_{L}^2\right)\right]~~. 
\end{equation*}
After the implementation of these rotations, the masses of the SM top quark $t$ (light top quark) and $T$ heavy top quark up to order $v^2/f^2$ can be expressed by~\cite{Buras15}  
\begin{equation} 
m_ {t}=\frac{\lambda_{1}\lambda_{2}v}{\sqrt{\lambda_{1}^{2}+\lambda_{2}^{2}}}%
\left[1+\frac{v^{2}}{f^{2}}\left(\frac{fv^{\prime}}{v^{2}}-\frac{1}{3}+
\frac{1}{2}\frac{\lambda_{1}^{2}}{\lambda_{1}^{2}+\lambda_{2}^{2}}\left( 1-\frac{\lambda_{1}^{2}}{\lambda_{1}^{2}+\lambda_{2}^{2}}\right)  \right)\right] 
\end{equation}
\begin{equation} 
m_ {T}=f\sqrt{\lambda_{1}^{2}+\lambda_{2}^{2}}\left[1+ \frac{v^{2}}{f^{2}}\frac{\lambda_{1}^2\lambda_{2}^2}{2(\lambda_{1}^2+\lambda_{2}^2)^2}\right]~~.  
\end{equation}

Since the SM top quark mass is known, the bound on $\lambda_{1}$ and $\lambda_{2}$  which is expected to be $\mathcal{O}(1)$ can be expressed as~\cite{Buras15}
\begin{equation}
\frac{1}{\lambda_{1}^2}+\frac{1}{\lambda_{2}^2}=\frac{v^2}{m_{t}^2}
\end{equation}
$\lambda_{1}$ and $\lambda_{2}$ can be written in terms of $m_{t}$ and $x_{L}$
\begin{equation}
\lambda_{1}=\frac{m_{t}}{v}\frac{1}{\sqrt{1-x_{L}}},~~~~~~~~\lambda_{2}=\frac{m_{t}}{v}\frac{1}{\sqrt{x_{L}}}
\end{equation}
where $x_{L}$ can vary in the range $0<x_{L}<1$.

\begin{figure}[t!]
\centering
\includegraphics[scale=0.6]{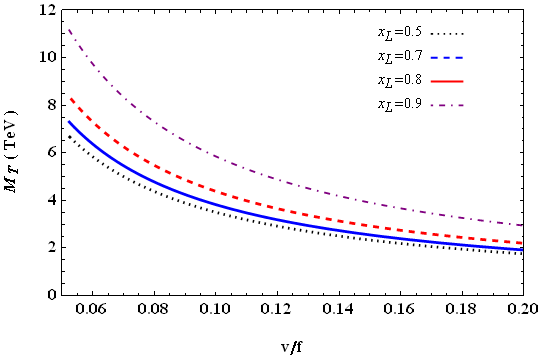}
\caption{The mass of heavy top quark, $T$, as a function of $v/f$ for different values of $x_{L}$.}
\label{fig:MT}
\end{figure}

As can be seen in Fig. 2, the masses of heavy top quark increase as $x_{L}$  increases but decrease as $v/f$ increases. To avoid the fine-tuning problems, the mass of the heavy top quark mass should be about 2 TeV ~\cite{Arkani-Hamed6,Yang32}

\section{The process $e^{-}e^{+}\longrightarrow Z_{H}\overline{T}t$ in the Littlest Higgs Model}
In this section, we study $e^{-}(p_{1})e^{+}(p_{2})\longrightarrow Z_{H}(p_{3})\overline{T}(p_{4})t(p_{5})$ in the LH model. The process gets additional contributions from $Z_{H}, Z_{L}, A_{H},A_{L}$. Relevant tree-level Feynman diagrams are shown in Fig. 3(a-h). In our numerical calculations, SM input parameters are taken as $M_{Z}$=91.2 GeV, $M_{W}$=80.2 GeV, $M_{H}$=125 GeV, $M_{t}$=173.07 GeV, $\alpha(m_{Z})$=1/128.8 and sin$^{2}{\theta_W}$=0.231. In addition to SM parameters, LH model includes four free parameters  $s$, $s^{\prime}$, $f$ and $x_{L}$ which depends on $\lambda_{1}$ and $\lambda_{2}$ as given in Eq. (2.17). Taking the constraints of the electroweak precision data into account (1 TeV $\leqslant$ $f$ $\leqslant$ 2 TeV, 0.75 $\leqslant$ $s$ $\leqslant$ 0.99, 0.6 $\leqslant$ $s^{\prime}$ $\leqslant$ 0.75), we take the parameters as $f$=1$\sim$2 TeV, $s/s^{\prime}$=0.8/0.6,0.7 which are consistent with the eletroweak precision data and $\lambda_{1}$ = $\lambda_{2}$. For the numerical calculations of the production cross section, all relevant vertices are implemented in the CalcHEP~\cite{Pukhov34}.
For Fig. 3(a-h), we present the amplitudes. Vector and axial vector  couplings for gauge bosons and fermion pairs are given in Table 2 where $x_{Z}^{W^{\prime}}$=$\frac{1}{2c_{w}}sc(c^2-s^2)$, $x_{Z}^{B^{\prime}}$=$\frac{5}{2s_{w}(s^{\prime}c^{\prime}({c^{\prime}}^{2}-{s^{\prime}}^{2}))}$, $\chi$=$-\frac{5}{6}-\frac{3}{2}{x_{L}}^{2}+2x_{L}$, $y_{e}$=$\frac{3}{5}$ and  $y_{u}$=$-\frac{2}{5}$. In our calculations we define the following notations for couplings:
\begin{equation}
\Lambda ^{V_{j}\overset{-}ff}=(g_{V}^{V_{j}\overset{-}ff}+g_{A}^{V_{j}\overset{-}ff}\gamma_5)
\end{equation}
where $j$=1, 2, 3, 4 correspond to $ A_{L}, Z_{L}, A_{H}, Z_{H}$, respectively. The amplitudes for Fig 3(a, b, c, d,e and h) are s channel processes with $q= p_{1}+p_{2}$. The amplitudes for the diagram Fig. 3(a-b) are written
\begin{equation*}\hspace{-2.6cm}
M_{a} =  \bar {u}(p_5) i\gamma_{\mu} \Lambda^{V_{1}\overline{t}t}\left[i\frac{(\slashed{q}^{\prime}+m_{t})}{q^{{\prime}^2}-m^{2}_{t}} \right] v(p_4)i\gamma_{\nu} \Lambda^{V_{4}\overline{T}t}\epsilon^{\nu}(p_3)\left[ -i\frac{g^{\mu\sigma}}{q^2}\right]\bar {v}(p_2)i\gamma_{\sigma}\Lambda ^{V_{1}\overset{-}ee}u(p_1)
\end{equation*}
\begin{equation*}\hspace{-0.4cm}
M_{b} = \sum_{j=2,3, 4} \bar {u}(p_5) i\gamma_{\mu} \Lambda ^{{V_j}\overline{t}t}\left[i\frac{(\slashed{q}^{\prime}+m_{t})}{q^{{\prime}^2}-m^{2}_{t}} \right] v(p_4)i\gamma_{\nu} \Lambda ^{V_{4}\overline{T}t}\epsilon^{\nu}(p_3)\left[ -i\frac{(g^{\mu\sigma}-q^{\mu}q^{\sigma}/m^2_{V_j})}{q^2-m^2_{V_j}+im_{V_j}\Gamma_{V_j}}\right] \bar {v}(p_2)i\gamma_{\sigma}\Lambda ^{V_{j}\overline{e}e}u(p_1)
\end{equation*}
where $q^{\prime}=q-p_{5}$.
For the diagrams Fig. 3 (c-d), the amplitudes are written

\begin{equation*}\hspace{-2.2cm}
M_{c} =\bar {u}(p_5) i\gamma_{\mu} \Lambda ^{{V_4}\overline{T}t}\epsilon^{\mu}(p_3)\left[i\frac{(\slashed{q}^{\prime}+m_{T})}{q^{{\prime}^2}-m^{2}_{T}} \right]v(p_4) i\gamma_{\nu} \Lambda ^{V_{1}\overline{T}T}\left[ -i\frac{g^{\nu\sigma}}{q^2}\right] \bar {v}(p_2)i\gamma_{\sigma}\Lambda ^{V_{1}\overline{e}e} u(p_1)
\end{equation*}
\begin{equation*}\hspace{-0.4cm}
M_{d} = \sum_{j=2,3} \bar {u}(p_5) i\gamma_{\mu} \Lambda ^{{V_4}\overline{t}T}\epsilon^{\mu}(p_3)\left[i\frac{(\slashed{q}^{\prime}+m_{T})}{q^{{\prime}^2}-m^{2}_{T}} \right]v(p_4) i\gamma_{\nu} \Lambda ^{V_{j}\overline{T}T}\left[ -i\frac{(g^{\nu\sigma}-q^{\nu}q^{\sigma}/m^2_{V_j})}{q^2-m^2_{V_j}+im_{V_j}\Gamma_{V_j}}\right] \bar {v}(p_2)i\gamma_{\sigma}\Lambda ^{V_{j}\overline{e}e}u(p_1)
\end{equation*}
where $q^{\prime}=q-p_{4}$. For the diagram Fig. 3e

\begin{equation*}\hspace{-0.4cm}
M_{e} = \sum_{j=2,3,4} \bar {u}(p_5) i\gamma_{\mu} \Lambda ^{{V_4}\overline{t}t}\epsilon^{\mu}(p_3)\left[i\frac{(\slashed{q}^{\prime}+m_{t})}{q^{{\prime}^2}-m^{2}_{t}} \right]v(p_4)i\gamma_{\nu} \Lambda ^{V_{j}\overline{T}t}\left[ -i\frac{(g^{\nu\sigma}-q^{\nu}q^{\sigma}/m^2_{V_j})}{q^2-m^2_{V_j}+im_{V_j}\Gamma_{V_j}}\right] \bar {v}(p_2)i\gamma_{\sigma}\Lambda ^{V_{j}\overline{e}e}u(p_1)
\end{equation*}
where $q^{\prime}=q-p_{4}$. The amplitudes for diagrams  Fig. 3f and 3g  are given by
\begin{equation*}\hspace{-0.4cm}
M_{f} = \sum_{j=2,3,4} \bar {u}(p_5) i\gamma_{\mu} \Lambda ^{{V_j}\overline{T}t}v(p_4)\left[ -i\frac{(g^{\mu\nu}-q^{\mu}q^{\nu}/m^2_{V_j})}{q^2-m^2_{V_j}+im_{V_j}\Gamma_{V_j}}\right]\bar {v}(p_2)i\gamma_{\sigma} \Lambda ^{V_{4}\overline{e}e}\epsilon^{\sigma}(p_3) 
\left[i\frac{(\slashed{q}^{\prime}+m_{e})}{q^{{\prime}^2}-m^{2}_{e}} \right]i\gamma_{\nu}\Lambda ^{V_{j}\overline{e}e}u(p_1)
\end{equation*}
\begin{equation*}\hspace{-0.4cm}
M_{g}= \sum_{j=2,3,4} \bar {u}(p_5) i\gamma_{\mu} \Lambda ^{{V_j}\overline{T}t}v(p_4)\left[ -i\frac{(g^{\mu\sigma}-q^{\mu}q^{\sigma}/m^2_{V_j})}{q^2-m^2_{V_j}+im_{V_j}\Gamma_{V_j}}\right]\bar {v}(p_2)i\gamma_{\sigma} \Lambda ^{V_{j}\overline{e}e} 
\left[i\frac{(\slashed{q}^{\prime}+m_{e})}{q^{{\prime}^2}-m^{2}_{e}} \right]
i\gamma_{\nu}\Lambda ^{V_{4}\overline{e}e}\epsilon^{\nu}(p_3)u(p_1)
\end{equation*}
where $q^{\prime}=p_{1}-q$ for the Fig. 3f and 3g. Fig. 3h  includes the coupling of fermion pairs to Higgs boson and two gauge bosons to Higgs boson. We denote the coupling of fermion pairs to Higgs boson as $\Lambda^{H\overset{-}{f}f}$ and the relevant coupling can be written as 

\begin{eqnarray}
\nonumber
\Lambda^{H\overset{-}{T}t} &=& -i\frac{\lambda_{1}^{2}}{\sqrt{%
\lambda_{1}^{2}+\lambda_{2}^{2}}}[1+\frac{v^{2}}{f^{2}}(-\frac{3}{2}-4\frac{f^{2}v^{^{\prime }2}}{v^{4}}%
+3\frac{fv^{\prime}}{v^{2}}+\frac{5}{2}\frac{\lambda_{1}^{2}}{\lambda_{1}^{2}+\lambda_{2}^{2}}%
\\&&%
-\frac{3}{2}\frac{\lambda^{4}}{(\lambda_{1}^{2}+\lambda_{2}^{2})^{2}})]P_{L}-%
\frac{i\lambda_{1}\lambda_{2}^{3}}{(\lambda_{1}^{2}+\lambda_{2}^{2})^{\frac{3}{2}}}%
\frac{v}{f}P_{R}\nonumber
\end{eqnarray}
The couplings of the gauge bosons to Higgs boson are given in Table 3. For Fig. 3h amplitude is given by
\begin{equation*}\hspace{-0.4cm}
M_{h} = \sum_{j=2,3,4} \bar {u}(p_5)\Lambda^{H\overline{T}t}v(p_4)\left[ \frac{i}{q^{{\prime}^2}-m^2_{H}}\right]\Lambda^{C_{k}}g_{{\mu}{\nu}}\epsilon^{\nu}(p_3) \left[ -i\frac{(g^{\mu\sigma}-q^{\mu}q^{\sigma}/m^2_{V_j})}{q^2-m^2_{V_j}+im_{V_j}\Gamma_{V_j}}\right]\bar{v}(p_2) 
i\gamma_{\sigma}\Lambda ^{V_{j}\overline{e}e}u(p_1)
\end{equation*}
where $q^{\prime}=p_{4}+p_{5}$.

\begin{figure}[t!]
\centering
\includegraphics[scale=0.4]{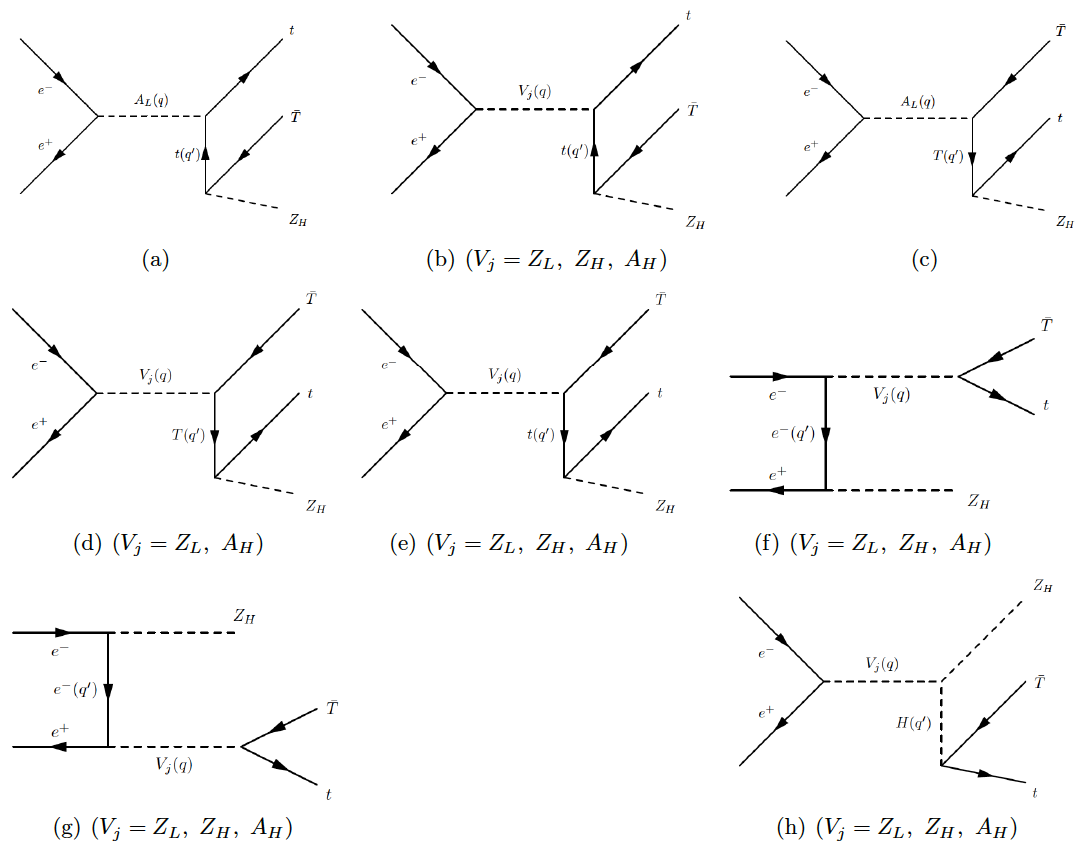}
\caption{Feynman diagrams of the process $e^{-}e^{+}\longrightarrow Z_{H}\overline{T}t$ in the Littlest Higgs Model.}
\label{fig:alldiagrams}  
\end{figure}

With the above amplitudes, we can calculate the production cross section. This process has contributions both s-channel and t-channel by exchancing of $A_{L},Z_{L},A_{H},Z_{H}$. Production cross section is plotted with respect to center of mass energy in Fig. 4 for different $s$ and $s^{\prime}$ for $f$=1 TeV and $\lambda_{1}$=$\lambda_{2}$ at $\sqrt{s}$=3000 GeV. Production cross section in s-channel is more dominant  than t-channel for the process. The values of the cross section are in the range $10^{-5}$ pb $\leqslant$ $\sigma$ $\leqslant$ 1.17 $10^{-4}$ pb in 2.83 TeV$\leqslant$ $\sqrt{s}$ $\leqslant$ 3 TeV ($s$/$s^{\prime}$=0.8/0.6) and 1.25 $10^{-5}$ pb $\leqslant$ $\sigma$ $\leqslant$ 7.65 $10^{-5}$ pb in 2.84 TeV$\leqslant$ $\sqrt{s}$ $\leqslant$ 3 TeV ($s$/$s^{\prime}$=0.8/0.7). If we take that integrated luminosity $L$=500 $fb^{-1}$, several tens of $Z_{H}\overline{T}t$ events will be generated.

\begin{figure}[t!]
\centering
\includegraphics[scale=0.9]{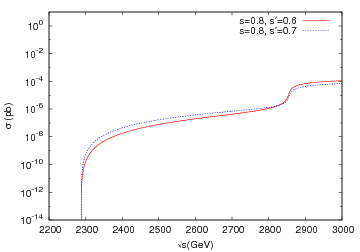}
\caption{The production cross section as a function of the center of mass energy for $f$=1TeV, $s/s^\prime$=0.8/0.6 (solid line) and $s/s^\prime$=0.8/0.7 (dot line) at $\sqrt{s}$=3000 GeV.}
\label{fig:1crs}  
\end{figure}

Fig. 5 shows the cross section as a function of the symmetry breaking scale $f$ for the different values of the mixing angles $s/s^\prime$ (0.8/0.6,0.7) at $\sqrt{s}$=3000 GeV. One can see from Fig. 5 that the production cross sections which are sensitive to the mixing parameters decrease as $f$ increases. The values of the cross section at $f$=1 TeV are  1.40 $10^{-5}$ pb and 1.24 $10^{-5}$ pb for $s$/$s^{\prime}$=0.8/0.6 and $s$/$s^{\prime}$=0.8/0.7, respectively. For the values of $f>$1060GeV, the cross sections are too small at $\sqrt{s}$=3000 GeV. Therefore heavy particles are difficult to be detected.   

\begin{figure}[t!]
\centering
\includegraphics[scale=0.9]{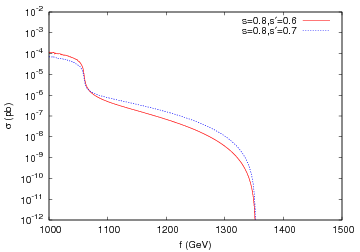}
\caption{The production cross section as a function of the symmetry breaking scale $f$ for $s/s^\prime$=0.8/0.6 (solid line) and $s/s^\prime$=0.8/0.7 (dot line) at $\sqrt{s}$=3000 GeV.}
\label{fig:f1}  
\end{figure}

\begin{figure}[t!]
\centering
\includegraphics[scale=0.9]{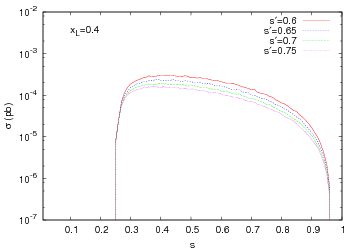}
\caption{The production cross section as a function of $s$ for $f$=1TeV, $x_{L}$=0.4  and $s^{\prime}$=0.6, 0.65, 0.7 and 0.75 at $\sqrt{s}$=3000 GeV.}
\label{fig:s1}
\end{figure}

\begin{figure}[t!]
\centering
\includegraphics[scale=0.8]{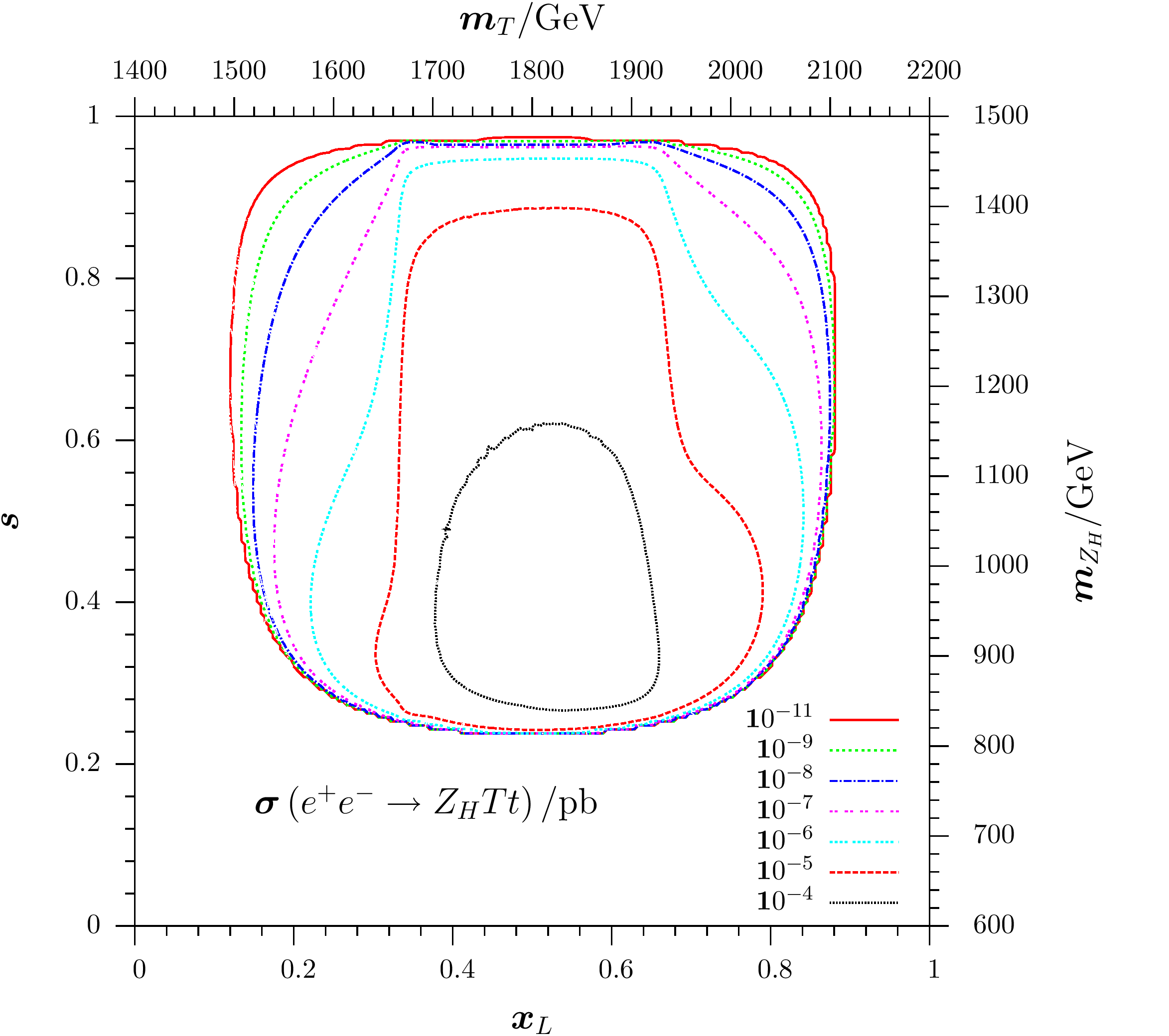}
\caption{Contours for the masses of the $Z_{H}$ and $T$ together with the mixing parameters at $\sqrt{s}$=3000 GeV and $f$= 1 TeV.}
\label{fig:contour_xL_s2}  
\end{figure}

Fig. 6 shows the cross section as a function of the mixing angle ($s$) for different values of $s^\prime$. We have taken $x_{L}$=0.4 and interval of the mixing angle $s^\prime$ considering the electroweak precision data for $f$=1 TeV. For 0.6 $\leqslant s^{\prime}\leqslant$ 0.75, the cross section decreases as $s^{\prime}$ increases. For $x_{L}$=0.4 and different values of the $s^{\prime}$, the cross section decreases as s increases. When we take 0.4 $\leqslant x_{L}\leqslant$ 0.6 the cross section increases as $x_{L}$ increases. For the values, out of 0.4 $\leqslant x_{L}\leqslant$ 0.6, the cross section is too small.

Fig.7 shows the constraints on heavy top quark and heavy Z boson together with the mixing parameters $s$ and $x_L$ at $\sqrt{s}$=3000 GeV. For 0.28 $\leqslant s \leqslant$ 0.62 and 0.38 $\leqslant x_{L}\leqslant$ 0.66, the production cross section is at order of $10^{-4}$ pb and the ranges of the masses are 1.7 TeV $\leqslant M_{T}\leqslant$ 1.9 TeV and  0.85 TeV $\leqslant M_{Z_H}\leqslant$ 1.16 TeV. We may exclude the region out of the values. For 0.24 $\leqslant s\leqslant$ 0.88 and 0.3 $\leqslant x_{L}\leqslant$ 0.78, the production cross section is at order of $10^{-5}$ pb and the ranges of mass are 1.64 TeV $\leqslant M_{T}\leqslant$ 2.04 TeV and 0.82 TeV $\leqslant M_{Z_H}\leqslant$ 1.4 TeV. For $Z_{H}$ and $T$ at $\sqrt{s}$=3000 GeV with $L$=500 $fb^{-1}$, a part of the mixing parameter space can be discovered at  $10^{-5}$ pb considering the electroweak precision data.
\begin{table}[hp!]
\centering
\scalebox{1.00}{
\begin{tabular}{|c|c|c|c|}
\hline 
 & Vertices &$g_{V_{i}}$& $g_{A_{i}}$ \\
\hline
1 & $A_{L}\overline{e}e$& -$eQ_{e}$ & 0 \\
\hline
2& $Z_{L}\overline{e}e$&
$\begin{array} {lcl} && -\frac{g}{2c_{W}}\{-\frac{1}{2}+2s_{w}^{2}-\frac{v^{2}}{f^{2}}[-\frac{c_{w}x_{Z}^{W^{\prime}}c}{2s}+
\\&&+ \frac{s_{w}x_{Z}^{B^{^{\prime }}}}{s^{^{\prime
}}c^{^{\prime }}}(2y_{e}-\frac{9}{5}+\frac{3}{2}c^{^{\prime }2})]\}\end{array}$&$\begin{array} {lcl} && -\frac{g}{2c_{W}}\{\frac{1}{2}-\frac{v^{2}}{f^{2}}[\frac{c_{w}x_{Z}^{W^{\prime}}c}{2s}+
\\&&+ \frac{s_{w}x_{Z}^{B^{^{\prime }}}}{s^{^{\prime
}}c^{^{\prime }}}(-\frac{1}{5}+\frac{1}{2}c^{^{\prime }2})]\} \end{array}$  \\   
\hline 
3 & $Z_{L}\overline{t}t$ &
$\begin{array} {lcl} && -\frac{g}{2c_{W}}\{\frac{1}{2}-\frac{4}{3}s_{w}^{2}-\frac{v^{2}}{f^{2}}%
[\frac{x_{L}^{2}}{2}+\frac{c_{w}x_{Z}^{W^{\prime}}c}{2s}+\\&&  \frac{s_{w}x_{Z}^{B^{^{\prime }}}}{s^{^{\prime%
}}c^{^{\prime }}}(2y_{u}+\frac{17}{15}-\frac{5}{6}c^{^{\prime}2}-\frac{1}{5}%
x_{L})]\}\end{array}$&%
$\begin{array} {lcl} && -\frac{g}{2c_{W}}\{-\frac{1}{2}-\frac{v^{2}}{f^{2}}%
[-\frac{x_{L}^{2}}{2}-\frac{c_{w}x_{Z}^{W^{\prime}}c}{2s}+\\&&\frac{s_{w}x_{Z}^{B^{^{\prime }}}}{s^{^{\prime%
}}c^{^{\prime }}}(\frac{1}{5}-\frac{1}{2}c^{^{\prime}2}-\frac{1}{5}%
x_{L})]\}\end{array}$\\
\hline
4 & $Z_{L}\overline{T}T$ &
$\begin{array} {lcl} && \frac{g}{2c_{W}}\{\frac{4}{3}s_{w}^{2}+\frac{v^{2}}{f^{2}}%
[-\frac{x_{L}^{2}}{2}+\\&&  \frac{s_{w}x_{Z}^{B^{^{\prime }}}}{s^{^{\prime%
}}c^{^{\prime }}}(2y_{u}+\frac{14}{15}-\frac{4}{3}c^{^{\prime}2}+\frac{1}{5}%
x_{L})]\}\end{array}$&%
$\begin{array} {lcl} && \frac{g}{2c_{W}}\frac{v^{2}}{f^{2}}\{\frac{x_{L}^{2}}{2}+\frac{s_{w}x_{Z}^{B^{^{\prime }}}}{s^{^{\prime%
}}c^{^{\prime }}}\frac{1}{5}x_{L}\}
\end{array}$\\
\hline
5 & $Z_{L}\overline{T}t$ &
$\begin{array} {lcl} && \frac{g}{2c_{W}}\{-\frac{vx_{L}}{2f}+\frac{v^{2}}{f^{2}}%
\frac{s_{w}x_{Z}^{B^{\prime}}}{s^{\prime}c^{\prime}}\frac{x_{L}\lambda_{2}}{5\lambda_{1}}%
+\frac{v^{3}}{f^{3}}(\frac{x_{L}^{3}}{4}\\&&-\frac{x_{L}}{2}\chi +x_{L}%
(\frac{c^{\prime}}{s^{\prime}}\frac{s_{w}x_{Z}^{B^{^{\prime }}}}{2}%
+\frac{c}{s}\frac{c_{w}x_{Z}^{W^{^{\prime }}}}{2}))\}\end{array}$&%
$\begin{array} {lcl} && \frac{g}{2c_{W}}\{\frac{vx_{L}}{2f}+\frac{v^{2}}{f^{2}}%
\frac{s_{w}x_{Z}^{B^{\prime}}}{s^{\prime}c^{\prime}}\frac{x_{L}\lambda_{2}}{5\lambda_{1}}%
+\frac{v^{3}}{f^{3}}(-\frac{x_{L}^{3}}{4}+\\&&\frac{x_{L}}{2}\chi -x_{L}%
(\frac{c^{\prime}}{s^{\prime}}\frac{s_{w}x_{Z}^{B^{^{\prime }}}}{2}%
+\frac{c}{s}\frac{c_{w}x_{Z}^{W^{^{\prime }}}}{2}))\}\end{array}$
\\
\hline

6 & $A_{H}\overline{e}e$ &
$\begin{array} {lcl} &&\frac{g^{\prime}}{2s^{\prime}c^{\prime}}(2y_{e}-\frac{9}{5}+%
\frac{3}{2}c^{^{\prime}2})\end{array}$&%

$\begin{array} {lcl} &&\frac{g^{\prime}}{2s^{\prime}c^{\prime}}(-\frac{1}{5}+%
\frac{1}{2}c^{^{\prime}2}) \end{array}$
\\
\hline
7 & $A_{H}\overline{t}t$ &$\begin{array} {lcl} &&\frac{g^{\prime}}{2s^{\prime}c^{\prime}}(2y_{u}+\frac{17}{15}-%
\frac{5}{6}c^{^{\prime}2}-\frac{1}{5}x_{L})\end{array}$

&$\begin{array} {lcl} &&\frac{g^{\prime}}{2s^{\prime}c^{\prime}}(\frac{1}{5}-%
\frac{1}{2}c^{^{\prime}2}-\frac{1}{5}x_{L})\end{array}$
\\
\hline
8 & $A_{H}\overline{T}t$ &$\begin{array} {lcl} &&\frac{g^{\prime}}{2s^{\prime}c^{\prime}}%
(\frac{1}{5}x_{L}\frac{\lambda_{2}}{\lambda_{1}}+%
\frac{v}{f}\frac{1}{2}c^{^{\prime}2}x_{L})\end{array}$

&$\begin{array} {lcl} &&\frac{g^{\prime}}{2s^{\prime}c^{\prime}}%
(\frac{1}{5}x_{L}\frac{\lambda_{2}}{\lambda_{1}}-%
\frac{v}{f}\frac{1}{2}c^{^{\prime}2}x_{L})\end{array}$
\\
\hline

9 & $A_{H}\overline{T}T$ &$\begin{array} {lcl} &&\frac{g^{\prime}}{2s^{\prime}c^{\prime}}(2y_{u}+\frac{14}{15}-%
\frac{4}{3}c^{^{\prime}2}+\frac{1}{5}x_{L})\end{array}$

&$\begin{array} {lcl} &&\frac{g^{\prime}}{2s^{\prime}c^{\prime}}\frac{1}{5}x_{L}\end{array}$
\\
\hline
10 & $Z_{H}\overline{e}e$ &$\begin{array} {lcl} &&-\frac{gc}{4s}\end{array}$

&$\begin{array} {lcl} &&\frac{gc}{4s}\end{array}$
\\
\hline
11 & $Z_{H}\overline{t}t$ &$\begin{array} {lcl} &&\frac{gc}{4s}\end{array}$

&$\begin{array} {lcl} &&-\frac{gc}{4s}\end{array}$
\\
\hline

12 & $Z_{H}\overline{T}t$ &$\begin{array} {lcl} &&\frac{gx_{L}vc}{4fs}\end{array}$

&$\begin{array} {lcl} &&-\frac{gx_{L}vc}{4fs}\end{array}$\\

\hline
\end{tabular}
}
\caption{Vector and axial couplings of fermions with light and heavy vector bosons\cite{Buras15}.}
\end{table}

\begin{table}[h!]
\centering
\begin{tabular}{|c|c|c|}
\hline 
$k$ & Vertices & $\Lambda^{C_{k}}g_{{\mu}{\nu}}$\\
\hline
1 & $Z_{H\mu}Z_{H\nu}H$ &
$\begin{array} {lcl} &&-\frac{i}{2}g^{2}vg_{\mu\nu} \end{array}$\\
\hline
2 & $Z_{L\mu}Z_{H\nu}H$&
$\begin{array} {lcl} && -\frac{i}{2}\frac{g^{2}}{c_{w}}\frac{(c^{2}-s^{2})}{2sc}%
vg_{\mu\nu}\end{array}$
\\
\hline
3 & $Z_{H\mu}A_{H\nu}H$&
$\begin{array} {lcl} && -\frac{i}{4}g g^{\prime}\frac{(c^{2}s^{^{\prime }2}+s^{2}c^{^{\prime }2})}%
{scs^{\prime}c^{\prime}}%
vg_{\mu\nu}\end{array}$
\\
\hline
\end{tabular}
\caption{The interaction vertices for $V_{i}V_{j}H$. $i$ and $j$ denote light ($L$) and heavy ($H$) gauge bosons~\cite{Han7}.}
\end{table}
\pagebreak
\pagebreak
\pagebreak

\section{Conclusion}
In this work, we calculate the associated production of heavy top quark ($T$) and heavy $Z$ boson ($Z_H$) in $e^{-}e^{+}$ colliders. It is found that with 500 $fb^{-1}$ of integrated luminosity, in a narrow range of the parameter space, $s/s^{\prime}$=0.8/0.6,0.7, 0.4 $\leqslant x_{L}\leqslant$ 0.6 and $f\lesssim$ 1060 GeV, a statical significance of 5$\sigma$ can be achieved. As it is seen in contours for the masses of $Z_H$ and $T$ for 0.24 $\leqslant s\leqslant$ 0.88 and 0.3 $\leqslant x_{L}\leqslant$ 0.77, the production cross section is at order of $10^{-5}$pb and the ranges of the masses are 1.64 TeV $\leqslant M_{T}\leqslant$ 2.04 TeV and 0.82 TeV $\leqslant M_{Z_H}\leqslant$ 1.4 TeV at $\sqrt{s}$=3000 GeV.

\section*{Acknowledgement}
The author would like to thank to T. M. Aliev and I. Turan for useful disscussions and quidance.

\end{document}